\documentclass[preprint,english,showpacs,11pt,floatfix]{revtex4}
\usepackage{babel,amsmath,amssymb,dcolumn}
\usepackage[dvips]{graphics}
\usepackage{hyperref}

\usepackage{amsfonts}
\usepackage{amssymb}
\usepackage[dvips]{graphicx}
\usepackage{latexsym}

\usepackage{dcolumn}
\usepackage{bm}

\begin{document}

\title{Thermodynamic Behavior of Friedmann Equation at
Apparent Horizon of FRW Universe}
\author{M. Akbar\footnote{Email address: akbar@itp.ac.cn} and Rong-Gen Cai\footnote{Email address: cairg@itp.ac.cn} }
\affiliation{
Institute of Theoretical Physics, Chinese Academy of Sciences, \\
P.O. Box 2735, Beijing 100080, China}

\begin{abstract}
It is shown that the differential form of Friedmann equation of a
FRW universe can be rewritten as the first law of thermodynamics $dE
= TdS + WdV$ at apparent horizon, where $E=\rho V$ is the total
energy of matter inside the apparent horizon, $V$ is the volume
inside the apparent horizon, $W=(\rho-P)/2 $ is the work density,
$\rho$ and $P$ are energy density and pressure of matter in the
universe, respectively. From the thermodynamic identity one can
derive that the apparent horizon $\tilde r_A$ has associated entropy
$S= A/4G$ and temperature $T = \kappa / 2\pi$ in Einstein general
relativity, where $A$ is the area of apparent horizon and $\kappa$
is the surface gravity at apparent horizon of FRW universe. We
extend our procedure to the Gauss-Bonnet gravity and more general
Lovelock gravity and show that the differential form of Friedmann
equations in these gravities can also be written as $dE = TdS + WdV$
at the apparent horizon of FRW universe with entropy $S$ being given
by expression previously known via black hole thermodynamics.
\end{abstract}

\maketitle

\section{Introduction}

 Semiclassical quantum properties
of black hole can be analyzed in the context of quantum field
theory in curved backgrounds, where matter is described by quantum
field theory while gravity enters as a classical background. In
this framework, it was discovered that black holes can emit
Hawking radiation with a temperature proportional to its surface
gravity at the black hole horizon and black hole has an entropy
proportional to its horizon area \cite{a1,a2}. The Hawking
temperature and horizon entropy together with the black hole mass
obey the first law of black hole thermodynamics \cite{a3}. The
gravitational entropy of black hole in Einstein gravity is given
by
\begin{equation}\label{1eq1}
S = \frac{A}{4G}
\end{equation}
where $A$ is the black hole horizon area and the units are such
that $c = \hbar = k = 1$. The Hawking temperature is given by
\begin{equation}\label{1eq2}
T_{H} = \frac{\kappa}{2\pi}
\end{equation}
where $\kappa$ is the surface gravity of the black hole. The
Hawking temperature together with black hole entropy is related by
the first law of black hole thermodynamics $TdS = dM$, where $M$
is the black hole mass (For a more general Kerr-Newman black hole,
the first law is $dM=TdS +\Omega dJ +\Phi dQ$). The black hole
thermodynamics and the statistical property of black hole entropy
have been investigated from many different points of view in
literature \cite{a4}. Nowadays it is widely believed that a black
hole behaves like an ordinary thermodynamic system and satisfies
laws of thermodynamics. If one identities the black hole mass $M$
as the energy $E$, obviously, a work term is absent in the first
law of black hole thermodynamics $dE = TdS$. To remedy this
drawback, More recently, Paranjape Sarkar and
Padmanabhan~\cite{PSP} have considered a special kind of
spherically symmetric black hole spacetimes, and found that it is
possible to interpret Einstein's equations as the thermodynamic
identity $TdS = dE + PdV$ by considering black hole horizon as the
system boundary. For related discussions on this issue see, for
example, references in \cite{Pad}.

 On the other hand, the
thermodynamical properties of the black hole horizon can be
generalized to the space-time horizons other than black hole
horizon. For example, the de Sitter space-time with radius $\ell$,
there is a cosmological event horizon. This horizon, like  black
hole horizon, can be regarded as a thermodynamical system \cite{a5}
associated with the Hawking temperature $T = 1 / 2\pi\ell$ and
entropy $S = A / 4G$, where $A = 4\pi\ell^{2}$ is the cosmological
horizon area of the de Sitter space-time. For an asymptotic de
Sitter space, like Schwarzschild-de Sitter space-time, there still
exists the cosmological horizon which behaves like a black hole
horizon with entropy proportional to the area of the cosmological
horizon and whose Hawking temperature is given by $T = \kappa /
2\pi$, where $\kappa$ is the surface gravity of the cosmological
horizon. It is easy to verify that the cosmological horizons of
these space-times satisfy the first law of black hole thermodynamics
of the form $TdS = -dM$ \cite{a6}, where the minus appears due to
the fact that when the black hole mass $M$ increases, the
cosmological horizon entropy decreases.

Indeed black hole physics implies that there is some relation
between the first law of thermodynamics and Einstein's equations.
Jacobson~\cite{a8} is the first one to seriously investigate such
a relation. Jacobson finds that it is indeed possible to derive
the Einstein's equations from the proportionality of entropy to
the horizon area together with the fundamental relation $\delta Q=
TdS$, assuming the relation holds for all local Rindler causal
horizons through each spacetime point. Here $\delta Q$ and $T$ are
the energy flux and Unruh temperature seen by an accelerated
observer just inside the horizon.

In Ref.~\cite{a11}, one of the present authors and Kim are able to
derive the Friedmann equations of an (n+1)-dimensional
Friedman-Robertson-Walker (FRW) universe with any spatial
curvature by applying the first law of thermodynamics ($TdS=-dE$)
to the apparent horizon of the FRW universe and by working out the
heat flow through the apparent horizon. In the process, an ansatz
is made: suppose that the apparent horizon has  temperature and
entropy expressed by
\begin{equation}\label{1eq3}
T = \frac{1}{2\pi \tilde{r}_{A}},~~~~~~ S = \frac{A}{4G}
\end{equation}
where $A$ is the area of the apparent horizon.  Also by using the
entropy expression of a static spherically symmetric black hole in
the Gauss-Bonnet gravity and in more general Lovelock gravity,
they reproduce the corresponding Friedmann equations in each
gravity.  The possible extensions to the scalar-tensor gravity and
$f(R)$ gravity theory have been studied in reference \cite{a12}.
In the cosmological setting, related discussions see also,
\cite{a7,a9,a10,ain}.

Note that in the process of deriving the Friedmann equations, $-dE$
is interpreted as the amount of energy flux crossing the apparent
horizon within an infinitesimal time interval $dt$ and the horizon
radius is assumed to be not changed during the internal. On the
other hand, we know that the Friedmann equations of FRW universe are
the field equations with a source of perfect fluid. In this
cosmological setup, different from the case of black hole spacetimes
discussed in \cite{PSP}, there is a well-defined concept of pressure
$P$ and energy density $\rho$. Therefore, it is very interesting to
see whether it is possible to rewrite the Friedmann equations  as a
thermodynamical identity $TdS = dE + PdV$ at apparent horizon of the
FRW universe? To resolve this issue we develop a procedure to study
the thermodynamical properties of Friedmann equations at apparent
horizon of a FRW universe and show that by employing Misner-Sharp
energy relation inside a sphere of radius $\tilde{r}_{A}$ of
apparent horizon, it is indeed possible to rewrite the differential
form of the Friedman equations as the form of the first law of
thermodynamic $dE = TdS + WdV$ with $W=(\rho-P)/2$, where $\rho$ and
$P$ are energy density and pressure of matter in the universe. In
the case of Einstein gravity, we find that the horizon entropy $S$
is proportional to the apparent horizon area and the temperature $T$
is given by $T = \kappa / 2\pi$ with surface gravity at the apparent
horizon, and the Misner-Sharp energy is just the total energy of
matter inside the apparent horizon $(E=\rho V)$. We extend this
procedure to the Gauss-Bonnet gravity and in more general Lovelock
gravity, and verify that the Friedmann equations at apparent horizon
can also be rewritten as a universal form $dE = TdS + WdV$ in these
gravities. It is important to mention here that one can pick up the
expressions for entropy $S$ from the identity $dE = TdS + WdV$
obtained from the field equations at apparent horizon, which agrees
with the expression previously derived by black hole thermodynamics.
In addition,  for the Gauss-Bonnet gravity and Lovelock gravity, the
energy in the first law is not the Misner-Sharp energy, but the
total energy ($\rho V$) of matter inside the apparent horizon.  This
will be seen shortly.

This paper is organized as follows. In Sec.~II, we shall summarize
the procedure constructed to study the thermodynamic properties of
the apparent horizon through the Friedmann equations of a FRW
universe. In Sec.~III, we shall apply our procedure to the field
equations of FRW universe in Einstein gravity. We shall extend our
procedure to the Gauss-Bonnet gravity and to more general Lovelock
gravity in Sec.~IV and Sec.~V, respectively. Finally in Sec.~VI,
we shall conclude our results with some discussions.


\section{The Procedure}

 We consider a spatially homogenous
and isotropic universe described by the FRW metric. The line
element of an $(n+1)$-dimensional FRW universe is represented by
\begin{equation}
\label{2eq1}
 ds^{2} = -dt^2 + a^2(t)\gamma_{ij}~ dx^{i} dx^{j},
\end{equation}
where the $n$-dimensional spatial hypersurfaces with negative,
zero or positive curvature are parameterized by $k = -1$, $0$ and
$1$, respectively, and $a(t)$ is the scale factor of the universe
with $t$ being the cosmic time. The metric $\gamma_{ij}$ is given
by
\begin{equation}
\label{2eq2} \gamma_{ij}dx^idx^j = \frac{dr^{2}}{1-k r^{2}} +
r^{2}d\Omega^{2}_{n-1}.
\end{equation}
Here $d\Omega^{2}_{n-1}$ is the metric of $(n-1)$-dimensional
sphere with unit radius and the spatial curvature constant $k =
1$, $0$ and $-1$ correspond to a closed, flat and open universe,
respectively. Using spherical symmetry, the metric (\ref{2eq1})
can be re-written as
\begin{equation}
 \label{2eq3}
  ds^{2} = h_{ab}dx^{a}dx^{b} +
\tilde{r}^{2}d\Omega^{2}_{n-1},
\end{equation}
where $\tilde{r}= a(t)r$ and $x^{0}=t$, $x^{1}=r$ and the two
dimensional metric $h_{ab} = {\rm diag}(-1, a^{2}/1-kr^{2})$. The
dynamical apparent horizon is determined by the relation
$h^{ab}\partial_{a}\tilde{r}\partial_{b}\tilde{r}=0$, which
implies that the vector $\nabla \tilde{r}$ is null on the apparent
horizon surface. The explicit evaluation of the apparent horizon
for the FRW universe gives the apparent horizon radius
\begin{equation}
\label{2eq4}
  \tilde{r}_{A} = 1/\sqrt{H^{2}+k/a^{2}}.
\end{equation}
The associated temperature $T = \kappa / 2\pi$ at the apparent
horizon is determined through the surface gravity
\begin{equation}\label{2eg5}
\kappa = \frac{1}{2\sqrt{-h}}
\partial_{a}(\sqrt{-h}h^{ab}\partial_{b}\tilde{r}).
\end{equation}
The explicit evolution of the surface gravity at apparent horizon
of FRW universe reads~\cite{a11}
\begin{equation}\label{2eq6}
\kappa = -\frac{1}{\tilde{r}_{A}}(1 -
\frac{\dot{\tilde{r}}_{A}}{2H\tilde{r}_{A}}),
\end{equation}
where an over-dot denotes the derivative with  respect to the
cosmic time $t$.
 We now introduce the total
energy $E$ inside a sphere of radius $\tilde{r}$ defined by
\begin{equation}
 \label{2eq7}
E = \frac{n(n-1)\Omega_{n}\tilde{r}^{n-2}}{16\pi G} (1 -
h^{ab}\partial_{a}\tilde{r}\partial_{b}\tilde{r}),
\end{equation}
where $V = \Omega_{n}\tilde{r}^{n}$ is the volume of an
$n$-dimensional sphere with radius $\tilde{r}$ and $\Omega_{n} =
\frac{\pi^{n/2}}{\Gamma(n/2 + 1)}$ being the volume of an
$n$-dimensional unit ball. The total energy (\ref{2eq7}) is actually
the direct $(n+1)$-dimensional generalization of the
(3+1)-dimensional one, given by Misner and Sharp~\cite{a23}. At the
apparent horizon $\tilde{r} = \tilde{r}_{A}$, the term
$h^{ab}\partial_{a}\tilde{r}\partial_{b}\tilde{r} = 0$ in equation
(\ref{2eq7}), therefore the total energy inside a sphere of radius
$\tilde{r}_{A}$ is given by
\begin{equation}\label{2eq8}
E = \frac{n(n-1)\Omega_{n}}{16\pi G}\tilde{r}_{A}^{n-2},
\end{equation}
which agrees with the expression for the mass in the
$(n+1)$-dimensional Schwarzschild black hole once the apparent
horizon is replaced by the event horizon of the black hole. We
consider the FRW universe as a thermodynamical system with apparent
horizon surface as a boundary of the system. In general the radius
of the apparent horizon $\tilde{r}_{A}$ is not constant but changes
with time. Let $d\tilde{r}_{A}$ be an infinitesimal change in radius
of the apparent horizon of FRW universe during a time of interval
$dt$. This small displacement $d\tilde{r}_{A}$ in the radius of
apparent horizon will cause a small change $dV$ in the volume $V$ of
the apparent horizon. This leads to build up two spherical systems
of space-time with radii $\tilde{r}_{A}$ and $\tilde{r}_{A} + d
\tilde{r}_{A}$ having a common source $T_{\mu\nu}$ of perfect fluid
with non-zero pressure $P$ and energy density $\rho$ near apparent
horizon. Each space-time describing a thermodynamical system and
satisfying Einstein equations, differs infinitesimally in the
extensive variables volume, energy and entropy by $dV$, $dE$ and
$dS$, respectively, while having same values the intensive variables
temperature $T$ and pressure $P$. Thus, for these two space-times
describing thermodynamical states, there must exist a certain
relation relating these thermodynamic quantities. It turns out it is
indeed the case. Mathematically, the main points of the procedure
can be summarized as follows.

i) Write down the Friedmann equation of FRW universe in term of
radius $\tilde{r}_{A}$ of the apparent horizon. Then taking its
differential, one gets a new form of Friedmann equation called the
differential form of the Friedmann equation. The differential form
of the field equations describes how the changes near apparent
horizon are related through the field equations.

ii) Multiply by a factor $n\Omega_{n}\tilde{r}_{A}(1 -
\frac{\dot{\tilde{r}}_{A}}{2H\tilde{r}_{A}})$ on both sides of the
differential form of the Friedmann equation and then simplify it to
get an equation of the form
\begin{equation}\label{2eq9}
TdS = n\Omega_{n} \tilde{r}_{A}^{n} (\rho + P) H (1 -
\frac{\dot{\tilde{r}}_{A}}{2H\tilde{r}_{A}}) dt,
\end{equation}
where $A = n\Omega_{n}\tilde{r}_{A}^{n-1}$ is the area of apparent
horizon, $H$ denotes the Hubble parameter, $T=\kappa/2\pi$,  and
$\rho$ and $P$ are energy density and pressure of perfect fluid in
the FRW universe, respectively.

iii)  Write down the expression for the total energy $E$ for
$(n+1)$-dimensional FRW universe inside a sphere of radius $
\tilde{r}_{A}$ and then find out $dE$, an infinitesimal change in
energy during a small interval $dt$ of time. After simplifying,
one may get a relation of the type
\begin{equation}\label{2eq10}
dE = n\Omega_{n}\tilde{r}_{A}^{n - 1}\rho d\tilde{r}_{A}  -
n\Omega_{n}\tilde{r}_{A}^{n}(\rho + P) H dt.
\end{equation}

iv) Using the relation $d\tilde r_A=-H\tilde r_A^3 (\dot
H-k/a^2)dt$~\cite{a11}, we find that equations (\ref{2eq9}) and
(\ref{2eq10}) result in  a thermodynamical identity
 \begin{equation}
 \label{2eq11}
dE = TdS + WdV,
\end{equation}
 where $W = \frac{1}{2} (\rho -
P)$. Here the quantity $W$ is nothing, but the work density
defined in \cite{a15} through $W =- \frac{1}{2} T^{ab}h_{ab}$.
Compared to the standard form of the first law of thermodynamics,
$dE=TdS-PdV$, the work density $W$ replaces the pressure $P$ in
our expression (\ref{2eq11}).

 In the present work, we shall apply this procedure to the Friedmann
equations in the Einstein gravity, Gauss-Bonnet gravity and
Lovelock gravity at apparent horizon of FRW universe and show that
the Friedmann equations in these gravities all can be rewritten as
a universal thermodynamical identity (\ref{2eq11}).


\section{Thermodynamic Behavior of Friedmann Equation in Einstein
Gravity}

 The Einstein field equations read
\begin{equation}\label{3eq1}
G_{\mu\nu} = 8 \pi G T_{\mu\nu},
\end{equation}
where $G_{\mu\nu}$ is the Einstein tensor and $T_{\mu\nu}$ is the
energy - momentum tensor of the matter fields. In the FRW
universe, due to the symmetries of the FRW metric, the
stress-energy tensor $T_{\mu\nu}$ must be diagonal, and by
isotropy the spatial components must be equal. The simplest
realization of such a stress-energy tensor is that of a perfect
fluid described by a time dependent energy density $\rho(t)$ and
pressure $P(t)$
\begin{equation}\label{3eq2}
T_{\mu\nu} = (\rho + P) U_{\mu} U_{\nu} + Pg_{\mu\nu},
\end{equation}
where $U^{\mu}$ is the four velocity of the  fluid. With the
conservation of stress-energy tensor $T^{\mu\nu}_{;\nu} = 0$, one
gets the continuity equation of the perfect fluid
\begin{equation}\label{3eq3}
\dot{\rho} + nH(\rho + P) = 0.
\end{equation}
Solving Einstein's equations (\ref{3eq1}) in the background of
metric (\ref{2eq1}) and assuming the energy-momentum tensor
(\ref{3eq2}) of perfect fluid, one gets the Friedmann equation
\begin{equation}\label{3eq4}
H^{2} + \frac{k}{a^{2}} = \frac{16 \pi G}{n(n-1)} \rho,
\end{equation}
where $H$ is the Hubble parameter with $H = \dot{a} / a$.
Combining (\ref{3eq3}) and (\ref{3eq4}), one has $\dot H
-\frac{k}{a^2} =-\frac{8\pi G}{n-1} (\rho +P)$. It can be seen
from the radius $\tilde{r}_{A} = 1 / \sqrt{H^{2} + k / a^{2}}$ of
the apparent horizon that in the case of a flat universe, i.e., $k
= 0$, the radius $\tilde{r}_{A}$ of the apparent horizon has the
same value as the radius $\tilde{r}_{H}$ of the Hubble horizon,
which is defined as the inverse of the Hubble parameter, that is,
$\tilde{r}_{H} = 1 / H$. On the other hand, the cosmological event
horizon defined by
\begin{equation}\label{3eq5}
\tilde{r}_{E} = a(t) \int_{t}^{\infty} \frac{dt}{a(t)},
\end{equation}
exists only for an accelerated expanding universe. As a
consequence, for a pure de Sitter universe with $k= 0$, the
apparent horizon, the Hubble horizon and the cosmological event
horizon have the same constant value $1 / H$. Note that though the
cosmological event horizon does not always exist for all FRW
universes, the apparent horizon and the Hubble horizon always do
exist. The apparent horizon has been argued to be a causal horizon
for a dynamical space time and is associated with gravitational
entropy and surface gravity \cite{a15,a14}. Thus for our purpose
it would be useful to study the thermodynamical properties of
Friedmann equations of FRW universe at the apparent horizon. (Note
that in some cases, thermodynamics is not well-defined for Hubble
horizon and event horizon~\cite{wang}). In terms of the apparent
horizon radius, the Friedmann equation (\ref{3eq4}) can be
rewritten as
\begin{equation}\label{3eq6}
\frac{1}{\tilde{r}_{A}^{2}} = \frac{16\pi G}{n(n-1)} \rho.
\end{equation}
Then by taking differential of equation (\ref{3eq6}) and using the
continuity equation (\ref{3eq3}), one gets the differential form
of the Friedmann equation
\begin{equation}\label{3eq7}
\frac{1}{\tilde{r}_{A}^{3}} d\tilde{r}_{A} = \frac{8\pi G}{n-1}
(\rho + P) Hdt.
\end{equation}
Multiplying both hand sides of equation (\ref{3eq7}) by a factor
$n\Omega_{n}\tilde{r}_{A}^{n}(1 -
\frac{\dot{\tilde{r}}_{A}}{2H\tilde{r}_{A}})$, one can rewrite this
equation in the form
\begin{equation}\label{3eq8}
\frac{\kappa}{2\pi}d(\frac{n\Omega_{n}\tilde{r}_{A}^{n-1}}{4 G}) =
-n\Omega_{n}\tilde{r}_{A}^{n} (\rho + P) H (1 -
\frac{\dot{\tilde{r}}_{A}}{2H\tilde{r}_{A}}) dt.
\end{equation}
From the left hand side of equation (\ref{3eq8}), one immediately
recognizes that the quantity $\frac{\kappa}{2\pi}$ and the
quantity $\frac{n\Omega_{n}\tilde{r}_{A}^{n-1}}{4 G}$ inside
parentheses on the left hand side are nothing, but the temperature
$T=\kappa/2\pi$ and entropy $S=A/4G$ ($A =
n\Omega_{n}\tilde{r}_{A}^{n-1}$ being the area of the apparent
horizon). Therefore the above equation can be rewritten as
\begin{equation}\label{3eq9}
TdS = -n\Omega_{n}\tilde{r}_{A}^{n} (\rho + P) H (1 -
\frac{\dot{\tilde{r}}_{A}}{2H\tilde{r}_{A}}) dt.
\end{equation}

Now we consider the Misner-Sharp energy (\ref{2eq8}) surrounded by
the apparent horizon $\tilde{r} = \tilde{r}_{A}$ of the FRW
universe, given by
\begin{equation}\label{3eq10}
E = \frac{n(n-1)\Omega_{n}}{16\pi G}\tilde{r}_{A}^{n-2}.
\end{equation}
Using equation (\ref{3eq6}), one gets
\begin{equation}\label{3eq11}
E = \Omega_{n}\tilde{r}_{A}^{n} \rho.
\end{equation}
This is nothing, but the total energy ($\rho V$) of matter inside
the apparent horizon. It means that in Einstein gravity, the
Misner-Sharp energy (\ref{2eq8}) surrounded by the apparent horizon
is just the total energy of matter inside the apparent horizon. It
no longer holds for Gauss-Bonnet gravity and Lovelock gravity.

 Taking differential of equation (\ref{3eq11}), we get
\begin{equation}\label{3eq12}
dE = n\Omega_{n}\tilde{r}_{A}^{n-1}\rho d\tilde{r}_{A} +
\Omega_{n}\tilde{r}_{A}^{n} d\rho.
\end{equation}
Substituting $\dot{\rho} = -nH(\rho + P)$ into (\ref{3eq12}), one
reaches
\begin{equation}\label{3eq13}
dE = n\Omega_{n}\tilde{r}_{A}^{n-1} \rho d\tilde{r}_{A} -
n\Omega_{n}\tilde{r}_{A}^{n}(\rho + P) H dt.
\end{equation}
With the help of (\ref{3eq13}), equation (\ref{3eq9}) can be
further rewritten to
\begin{equation}\label{3eq14}
dE = TdS + \frac{1}{2}n\Omega_{n}\tilde{r}_{A}^{n-1}(\rho -
P)d\tilde r_A.
\end{equation}
Note that the volume $V= \Omega_{n}\tilde{r}_{A}^{n}$ and let
$W\equiv (\rho-P)/2$. The above equation can be finally written to
the form
\begin{equation}
\label{3eq15}
dE = TdS + WdV.
\end{equation}
Comparing this with the standard form of the first law of
thermodynamics, the negative pressure term is replaced by the term
$W$. In fact, this is not strange, the result (\ref{3eq15}) is
nothing, but the expression of the unified first law \cite{a15} of
thermodynamics in the setup of FRW universe. In the second
reference in \cite{a15}, the author derived a similar formula for
the trapping horizon of dynamical black hole.  Here, it is
important to note that the thermodynamic identity (\ref{3eq15}) is
obtained by using the Friedmann equation of FRW universe together
with the feature of apparent horizon.

In conclusion, by applying the Misner-Sharp energy relation,  the
Friedmann equation of FRW universe can be expressed as a
thermodynamical identity $dE = TdS + WdV$ at the apparent horizon.
On the other hand, from the relation (\ref{3eq15}) we may
``derive" that the apparent horizon has an associated
thermodynamics with temperature $T = \kappa/ 2\pi$ and entropy
$S=A/4G$, where $\kappa$ is the surface gravity and $A$ is the
area of apparent horizon.


\section{Thermodynamic Behavior of Friedmann equation in
Gauss-Bonnet Gravity}

In the previous section, we have studied the behavior of Friedmann
equation at apparent horizon of FRW universe in Einstein gravity
 and identified its behavior as a thermodynamical system satisfying
the unified first law of thermodynamics of the form $dE =TdS +
WdV$. It has been found that the apparent horizon of FRW universe
has an entropy proportional to its horizon area, very like the
black hole horizon entropy obeying the so-called area formula
\cite{a16}. However, it is well known that the area formula of
black hole entropy no longer holds in higher derivative gravities.
So it would be interesting to see whether, one can identify or not
the Friedmann equation as a thermodynamical system near apparent
horizon of FRW universe with entropy of expression with structure
similar to the black hole horizon entropy in these gravities by
the procedure developed in the previous section. In this section,
we shall continue the previous procedure for a special form of
higher derivative gravity, called Gauss-Bonnet gravity. This
theory contains a special combination of curvature-squared term,
added to the Einstein-Hilbert action. The Gauss-Bonnet term is
given by
\begin{equation}\label{4eq1}
R_{GB} = R^{2} - 4R_{\mu\nu}R^{\mu\nu} +
R_{\mu\nu\gamma\delta}R^{\mu\nu\gamma\delta}.
\end{equation}
The Gauss-Bonnet term naturally appears in the low energy
effective action of heterotic string theory. The Gauss-Bonnet term
is a topological term in four dimensions, and thus does not have
any dynamical effect in these dimensions. The action of the
Gauss-Bonnet gravity can be written by
\begin{equation}\label{4eq2}
S = \frac{1}{16\pi G}\int d^{n+1}x \sqrt{-g} (R + \alpha R_{GB}) +
S_{m},
\end{equation}
where $R$ is the $(n+1)$-dimensional Ricci scalar, $S_{m}$ is the
action of matter and $\alpha$ is a constant with the dimension
$(length)^{2}$. In the case of superstring theory in low energy
limit $\alpha$ is related to the inverse string tension and is
positive definite. The Gauss-Bonnet action is a natural extension of
the Einstein theory in the sense that no derivatives higher than
second order appear in the field equations. The equations of motion
for the action (\ref{4eq2}) are given by
\begin{equation}\label{4eq3}
G_{\mu\nu} + \alpha H_{\mu\nu} = 8\pi G T_{\mu\nu},
\end{equation}
where $ G_{\mu\nu} = R_{\mu\nu} - \frac{1}{2}g_{\mu\nu} R$, and
\begin{equation}\label{4eq4}
H_{\mu\nu} = 2(R R_{\mu\nu} - 2 R_{\mu\lambda}R^{\lambda}_{\nu} -
2 R^{\gamma\delta}R_{\gamma\mu\delta\nu} +
R_{\mu}^{\alpha\gamma\delta}R_{\alpha\nu\gamma\delta}) -
\frac{1}{2}g_{\mu\nu}R_{GB}.
\end{equation}
In the vacuum Gauss-Bonnet gravity with/without a cosmological
constant, static black hole solutions have been found and the
associated thermodynamics has been discussed \cite{ain2,a17}. In
this theory, the static, spherically symmetric black hole has the
form
\begin{equation}\label{4eq5}
ds^{2} = - e^{\lambda(r)}dt^{2} + e^{\nu(r)}dr^{2} + r^{2}
d\Omega_{n-1}^{2},
\end{equation}
with
\begin{equation}\label{4eq6}
e^{\lambda(r)} = e^{- \nu(r)} = 1 + \frac{r^{2}}{2\tilde
{\alpha}}\left (1 - \sqrt{1 + \frac{64\pi G
\tilde{\alpha}M}{n(n-1)\Omega_{n} r^{n}}}\right),
\end{equation}
where $\tilde {\alpha} = (n-2)(n-3)\alpha$ and $M$ is the mass of
black hole. In the limit $\alpha \rightarrow 0$, the above metric
reduces to the Schwarzschild metric in Einstein gravity. The
entropy of the black hole has the following form \cite{a17}
\begin{equation}\label{4eq7}
S = \frac{A}{4 G}\left(1 + \frac{n-1}{n-3} \frac{2
\tilde{\alpha}}{r_{+}^{2}}\right ),
\end{equation}
where $A = n\Omega_{n}r_{+}^{n-1}$ is the horizon area and $r_{+}$
is the horizon radius of the black hole. The authors of
reference~\cite{a11} have assumed that the entropy formula
(\ref{4eq7}) also holds for the apparent horizon of FRW universe
in the Gauss-bonnet gravity and the apparent horizon has the same
expression for entropy but replacing the black hole horizon radius
$r_{+}$ by the apparent horizon radius $\tilde{r}_{A}$, i.e.,
\begin{equation}\label{4eq8}
S = \frac{A}{4 G}\left(1 + \frac{n-1}{n-3} \frac{2
\bar{\alpha}}{\tilde{r}_{A}^{2}}\right ),
\end{equation}
with $A = n\Omega_{n}\tilde{r}_{A}^{n-1}$ being the area of the
apparent horizon. They reproduced Friedmann equations by applying
the relation $TdS = \delta Q$ to the apparent horizon with the
assumption that the apparent horizon still has the horizon
temperature $T = \frac{1}{2\pi\tilde{r}_{A}}$. Now we show that
the Friedmann equation of a FRW universe in the Gauss-Bonnet
gravity also can be rewritten as the thermodynamic identity $dE =
TdS +WdV$ at the apparent horizon.

The Friedmann equation for a FRW universe with perfect fluid as
its source in the Gauss-Bonnet gravity is \cite{a19}
\begin{equation}\label{4eq9}
(H^{2} + \frac{k}{a^{2}}) + \tilde{\alpha} (H^{2} +
\frac{k}{a^{2}})^{2} =  \frac{16\pi G}{n(n-1)} \rho.
\end{equation}
In terms of the apparent horizon radius, the Friedmann equation
can be written as
\begin{equation}\label{4eq10}
\frac{1}{\tilde{r}_{A}^{2}} + \tilde{\alpha}
\frac{1}{\tilde{r}_{A}^{4}} =  \frac{16\pi G}{n(n-1)} \rho
\end{equation}
According to our procedure, one gets the differential form of the
equation of motion by taking differential of equation
(\ref{4eq10})
\begin{equation}\label{4eq11}
\frac{1}{\tilde{r}_{A}^{3}} d\tilde{r}_{A} + 2 \tilde{\alpha}
\frac{1}{\tilde{r}_{A}^{5}} d\tilde{r}_{A} =  \frac{8\pi G}{(n-1)}
(\rho + P) Hdt,
\end{equation}
where the continuity equation, $\dot \rho = -n H (\rho +P)$, has
been used. Now we multiply both hand sides of the above equation
again by a factor $n\Omega_{n}\tilde{r}_{A}^{n}(1 -
\frac{\dot{\tilde{r}}_{A}}{2H\tilde{r}_{A}})$ and rewrite the above
equation to the form
\begin{equation}\label{4eq12}
\frac{\kappa}{2\pi} d\left ((\frac{n\Omega_{n}\tilde{r}_{A}^{n-1}}{4
G})(1 + \frac{n-1}{n-3}
\frac{2\tilde{\alpha}}{\tilde{r}_{A}^{2}})\right ) = -
n\Omega_{n}\tilde{r}_{A}^{n} (\tilde{\rho} + \tilde{P}) H (1 -
\frac{\dot{\tilde{r}}_{A}}{2H\tilde{r}_{A}}) dt.
\end{equation}
The first term in the left hand side of this equation
(\ref{4eq12}) is in the form $TdS$. If we take
\begin{eqnarray}\label{4eq13}
S &=& \frac{n\Omega_{n}\tilde{r}_{A}^{n-1}}{4 G}(1 +
\frac{n-1}{n-3} \frac{2 \bar{\alpha}}{\tilde{r}_{A}^{2}}),
\nonumber \\
T &=& \frac{\kappa}{2\pi},
\end{eqnarray}
then the equation (\ref{4eq12}) can be rewritten as
\begin{equation}\label{4eq14}
TdS = -n\Omega_{n}\tilde{r}_{A}^{n} (\tilde{\rho} + \tilde{P}) H(1 -
\frac{\dot{\tilde{r}}_{A}}{2H\tilde{r}_{A}})dt.
\end{equation}
On the other hand, we note that the right hand side of the above
equation is the same as the case in Einstein gravity (\ref{3eq9}).
Thus one can immediately rewrite (\ref{4eq14}) as  the
thermodynamical identity
\begin{equation}\label{4eq15}
dE = TdS + WdV.
\end{equation}
Here we would like to remind the readers that in equation
(\ref{4eq15}), the energy $E=\rho V$ is the total energy of matter
(\ref{3eq11}) inside the apparent horizon,  not the Misner and Sharp
energy inside the apparent horizon. The reason is that the Misner
and Sharp energy (\ref{3eq10}) cannot be written to the form
(\ref{3eq11}) in the case of the Gauss-Bonnet gravity.

Thus, once again, we are able to express the Friedmann equation in
the Gauss-Bonnet gravity as a thermodynamical identity
(\ref{4eq15}), here the temperature and entropy associated with the
apparent horizon are given by (\ref{4eq13}).


\section{Thermodynamic Behavior of Friedmann equations in Lovelock
Gravity}

In this section we extend the previous discussions to the more
general Lovelock gravity. The Lovelock theory of gravity generalizes
Einstein gravity when space time has a dimension greater than four.
In this case the most general Lagrangian \cite{a20} that gives
second order equations for the metric, is the sum over the
dimensionally extended Euler densities
\begin{equation}\label{5eq1}
L = \sum_{n = 0}^{m} c_{n}L_{n},
\end{equation}
where $c_{n}$ is an arbitrary constant and $L_{n}$ is the Euler
density of a 2n-dimensional manifold
\begin{equation}\label{5eq2}
L_{n} = 2^{-n}\delta_{c_{1}d_{}\cdots
c_{n}d_{n}}^{a_{1}b_{1}\cdots
a_{n}b_{n}}R_{a_{1}b_{1}}^{c_{1}d_{1}} \cdots
R_{a_{n}b_{n}}^{c_{n}d_{n}}
\end{equation}
where the generalized delta function
 $\delta_{c_{1}d_{} \cdots c_{n}d_{n}}^{a_{1}b_{1} \cdots a_{n}b_{n}}$
is totally antisymmetric in both sets of indices and
$R_{\alpha\beta}^{\gamma\delta}$ are the components of the curvature
tensor. $L_{0}$ is set to one, therefore, the constant $c_{0}$ is
just the cosmological constant. $L_{1}$ gives us the usual curvature
scalar term. In order for the general relativity to be recovered in
the low energy limit, the constant $c_{1}$ has to be positive. For
simplicity, we can set $c_{1} = 1$. $L_{2}$ is just the Gauss-Bonnet
term. Although the Lagrangian of the Lovelock gravity contains
higher order derivatives curvature terms, there are no terms with
more than second order derivatives of metric in equations of motion
just as in Gauss-Bonnet gravity. Therefore, in this sense, the
Lovelock gravity theory is not a higher derivative gravity theory.
Lovelock field equations are much more complicated than Einstein's
equations, but they can still be solved for some simple models as
Friedmann universe and spherically symmetric black holes.  The
static spherically symmetric black hole solutions can be obtained in
this theory in the sense that the metric function is determined by
solving for a real root of a polynomial equation \cite{ain2}. More
recently, topological black hole solutions have been also found in
the Lovelock gravity \cite{a22} (see also \cite{others}). The
horizon of these black holes can be hypersurface with a positive,
zero or negative constant scalar curvature. In particular, it has
been shown that the entropy of black hole horizon has a simple
expression in terms of the horizon radius, while the expression for
the metric function and causal structure of these black holes could
be quite involved. For an $(n+1)$-dimensional static, spherically
symmetric back hole with metric
\begin{equation}\label{5eq3}
ds^{2} = - f(r) dt^{2} + f^{-1}(r) dr^{2} + r^{2}
d\Omega_{n-1}^{2},
\end{equation}
the metric function is given by $f(r) = 1 - r^{2}F(r)$, where
$F(r)$ is determined by solving for real roots of the following
$m$th-order polynomial equation
\begin{equation}\label{5eq4}
\sum_{i = 0}^{m} \hat{c}_{i} F^{i}(r) = \frac{16\pi G M}{n(n-1)
\Omega_{n}r^{n}}.
\end{equation}
Here, $M$ is a constant of integration, which is just the mass of
the black hole, and the coefficients $\hat{c}_{i}$ are given by
\begin{equation}\label{5eq5}
\hat{c}_{0} = \frac{c_{0}}{n(n-1)}, ~~~ \hat{c}_{1} = 1, ~~~~
\hat{c}_{i} = c_{i} \prod_{j = 3}^{2m} (n + 1 - j) ~~~ {\rm for}
~~~ i > 1.
\end{equation}
The black hole entropy in terms of the horizon radius $r_{+}$ can
be expressed as \cite{a22}
\begin{equation}\label{5eq6}
S = \frac{A}{4G}\sum_{i = 1}^{m} \frac{i(n-1)}{n - 2i + 1}
\hat{c}_{i} r_{+}^{2 - 2i},
\end{equation}
where $A = n\Omega_{n}r_{+}^{n - 1}$ is the horizon area of the
black hole. The above expression of black hole entropy does not
contain the cosmological constant term $c_{0}$ because the black
hole entropy depends only upon the horizon geometry of the black
hole. The entropy formula (\ref{5eq6}) of black hole also holds
for the apparent horizon of FRW universe and the apparent horizon
has the same expression for entropy in the Lovelock gravity but
the black hole horizon radius $r_{+}$ is replaced by the apparent
radius $\tilde{r}_{A}$~\cite{a11}. That is, the apparent horizon
has the entropy
\begin{equation}\label{5eq7}
S = \frac{A}{4G}\sum_{i = 1}^{m} \frac{i(n-1)}{n - 2i + 1}
\hat{c}_{i} \tilde{r}_{A}^{2 - 2i}.
\end{equation}
The Friedmann equation of a FRW universe in the Lovelock gravity
is~\cite{a11}
\begin{equation}\label{5eq8}
\sum_{i = 1}^{m}  \hat{c}_{i}(H^{2} + \frac{k}{a^{2}})^{i} =
\frac{16\pi G}{n(n - 1)} \rho.
\end{equation}
In terms of the apparent horizon radius $\tilde{r}^2_{A} =
1/(H^{2}+k/a^{2})$, the Friedmann equation (\ref{5eq8}) can be
rewritten as
\begin{equation}\label{5eq9}
\sum_{i = 1}^{m}  \hat{c}_{i}(\tilde{r}_{A})^{-2i} = \frac{16\pi
G}{n(n - 1)} \rho.
\end{equation}
One can get the differential form of the equation by taking
differential of equation (\ref{5eq9}) and then using the
continuity equation
\begin{equation}\label{5eq10}
\sum_{i = 1}^{m} i \hat{c}_{i}(\tilde{r}_{A})^{-2i-1}
d\tilde{r}_{A} = \frac{8\pi G}{(n - 1)} (\rho + P) Hdt.
\end{equation}
We again multiply both hand sides of the above equation by
$n\Omega_{n}\tilde{r}_{A}^{n}(1 -
\frac{\dot{\tilde{r}}_{A}}{2H\tilde{r}_{A}})$. And then arrange the
left hand side,  we have
\begin{equation}\label{5eq11}
\frac{\kappa}{2\pi}  d\left
(\frac{n\Omega_{n}\tilde{r}_{A}^{n-1}}{4G} \sum_{i = 1}^{m}
\frac{i(n-1)}{n - 2i + 1} \tilde{c}_{i} \tilde{r}_{A}^{2-2i} \right
) = -n\Omega_{n}\tilde{r}_{A}^{n} (\rho + P) H(1 -
\frac{\dot{\tilde{r}}_{A}}{2H\tilde{r}_{A}}) dt.
\end{equation}
The left hand side of the above equation is of the form $TdS$.
Thus the above equation can be rewritten as
\begin{equation}\label{5eq12}
TdS = -n\Omega_{n}\tilde{r}_{A}^{n} (\rho + P) H(1 -
\frac{\dot{\tilde{r}}_{A}}{2H\tilde{r}_{A}})dt.
\end{equation}
Once again, the right hand side of equation (\ref{5eq12}) has the
same form as the case in Einstein gravity. Therefore we can
finally rewrite the Freidmann equation into the universal form
\begin{equation}\label{5eq13}
dE = TdS + WdV,
\end{equation}
for the more general Lovelock gravity.  Once again, here the energy
$E$ is the total energy of matter inside the apparent horizon, not
the Misner-Sharp energy, as the case of the Gauss-Bonnet gravity.


\section{Conclusion and Discussion}

In this work we have shown that the differential form of Friedmann
equation can be rewritten as a  form, $dE = TdS + WdV$, at the
apparent horizon of a FRW universe with any spatial curvature in
arbitrary dimensions. Here $E$ the total energy ($\rho V$) of matter
inside the apparent horizon, $W=(\rho-P)/2$ and $V$ is the volume
inside the apparent horizon. Compared to the standard form of the
first law of thermodynamics, the negative pressure term $-P$ is
replaced by the work density $W$. Note that for pure de Sitter
spacetime, $\rho=-P$, one then has a standard form $dE =TdS-PdV$. We
have also shown that the Friedmann equations in the Gauss-Bonnet
gravity and Lovelock gravity can also be expressed as the universal
form. In particular, if associate a temperature $T = \kappa / 2\pi$
to the apparent horizon, we can obtain an associated entropy
$S=A/4G$ with the apparent horizon in the Einstein gravity, which
has the same form as that of black hole entropy. In the Gauss-Bonnet
gravity and Lovelock gravity, we have also obtained corresponding
expressions of entropy, they keep the same forms as those of black
hole entropy in each gravity. In other words, if we regard that the
apparent horizon has a universal temperature $T = \kappa / 2\pi$, we
can pick up in our procedure the expression of entropy in different
gravity theories. The resulting expressions of entropy have the same
forms as obtained previously by using black hole thermodynamics. In
addition, let us mention that in Einstein gravity, the total energy
$E$ of matter inside the apparent horizon is just the Misner-Sharp
energy, but they are not equal in Gauss-Bonnet gravity and Lovelock
gravity.

Here more remarks are in order. First, it can be seen from
(\ref{2eq6}) that if $\dot {\tilde r}_A <2H\tilde r_A$, the apparent
horizon has a negative surface gravity; if one further defines
temperature $T=\kappa/2\pi$, the temperature is negative! This case
is quite similar to the case of the cosmological event horizon in
the Schwarzschild-de Sitter spacetime. In that case, one should
define temperature $T=|\kappa |/2\pi$, and when the energy $E$
increases inside the apparent horizon, the apparent horizon radius
$\tilde r_A$ decreases. The universal form should change to $-dE =
TdS +W|dV|$ in this case. In addition, like the case of black hole
spacetime, the temperature defined in this way only depends on the
geometry, but not gravity theory under study. For the FRW universe,
the apparent horizon has a universal expression $T=\kappa/2\pi$ with
$\kappa$ given by (\ref{2eq6}).

 Second, in \cite{a11}, Cai and Kim have derived the Friedmann equations
 by applying the first law of thermodynamics, $TdS=-dE$, to the
 apparent horizon of a FRW universe with the assumption that the
 apparent horizon has temperature $T=1/2\pi \tilde r_A$ and
 entropy $S=A/4G$. One might worry that the result in \cite{a11} is
 not consistent with the one in the present paper. This is not the case,
 in fact, they are consistent with each other.  To be not
 confused, first,
 we would like to stress here that the notation $dE$ in \cite{a11}
 is quite different from the same one used in the present manuscript.
 In \cite{a11}, $-dE$ is actually just the heat flux $\delta Q$ in
 \cite{a8} crossing the apparent horizon within an infinitesimal internal
 of time $dt$. The quantity is given by
\begin{equation}\label{6eq1}
\delta Q=-dE = n\Omega_{n}\tilde{r}_{A}^{n}(\rho + P) Hdt.
\end{equation}
In this calculation, the apparent horizon radius has been assumed to
be fixed.  In this manuscript we have used the matter energy $E$
given in (\ref{3eq11}) inside the apparent horizon.   We have
assumed that $d\tilde{r}_{A}$ be the infinitesimal change in the
radius of the apparent horizon in a small interval of time $dt$
which causes a small change $dV$ in volume of the apparent horizon.
Since the matter energy $E$ is directly related with the radius of
the apparent horizon, therefore, the change of apparent horizon
radius will change the energy $dE$ inside the apparent horizon.  By
this procedure, we worked out the change of energy $dE$ inside the
apparent horizon is
\begin{equation}\label{6eq2}
dE = n\Omega_{n}\tilde{r}_{A}^{n-1} \rho d\tilde{r}_{A} -
n\Omega_{n}\tilde{r}_{A}^{n}(\rho + P) Hdt.
\end{equation}
Therefore in our definition, a new term
$n\Omega_{n}\tilde{r}_{A}^{n-1} \rho d\tilde{r}_{A}$ appears.
Since the apparent horizon radius is assumed to be fixed in
calculating (\ref{6eq1}) in \cite{a11} (see also
\cite{a7,a9,a10}), from (\ref{2eq6}) the temperature therefore has
the form $T=1/2\pi \tilde r_A$ in that case.
 Furthermore, a natural consequence is that  the term of volume change
 is absent in \cite{a11}.

 Third, one interesting question may arise; whether one can always express the
Friedmann equations to the thermodynamic identity $TdS = dE + WdV$
at apparent horizon in any gravity theory? Since in Einstein,
Gauss-Bonnet and more general Lovelock gravity theories, the fact
that the Friedmann equations can be rewritten to the universal form,
$dE = TdS + WdV$, might be related to the observation that in these
gravities, the equations of field can be derived from a holographic
surface term~\cite{Padd}. If it is possible, is it always possible
to pick up the expressions for entropy  from the identity $dE = TdS
+ WdV$, which agree with previously known results? For example, in
cases of $f(R)$ and scalar tensor gravities, can one apply this
approach to identify field equations as the universal thermodynamic
form at apparent horizon and pick up the expressions for entropies
in these gravities?  Partial results of these issues are obtained
recently~\cite{ACC}.

Finally,  one more question is whether this is a procedure
applicable for all types of horizons of space-times?  These issue
are certainly associated with the holographic properties of
gravity. It would be of great interest to examine further the
consequences of these observations to the holographic principle.

\section*{Acknowledgments}
The authors thank  Y.G. Gong and T. Padmanabhan  for helpful
correspondences. The work was supported in part by a grant from
Chinese Academy of Sciences, by NSFC under grants No. 10325525 and
No. 90403029.



\begin{thebibliography}{99}

\bibitem{a1} S. W. Hawking, Commun. Math. Phys. {\bf{43}}, 199
(1975).
\bibitem{a2} J. D. Bekenstein, Phys. Rev. D\textbf{7}, 2333
(1973).
\bibitem{a3} J. M. Bardeen, B. Carter and S. W. Hawking, Commun. Math. Phys. \textbf{31}, 161
(1973).
\bibitem{a4} J. Bekenstein, ``Do we understand black hole entropy"? arXiv: gr-qc/9409015;
U.H. Gerlach, Phys. Rev. D\textbf{15}, 1479 (1976); G. t'Hooft,
Nucl. Phys. \textbf{B256}, 727 (1985); J. York, Phys. Rev.
D\textbf{15}, 2929 (1985); W.H. Zurek and K.S. Thorne, Phys. Rev.
Lett. \textbf{54}, 2171 (1985);  L. Bombelli et al., Phys. Rev.
D\textbf{34},373(1986); R. D. Sorkin, arXiv: gr-qc/9705006.

\bibitem{PSP}A.~Paranjape, S.~Sarkar and T.~Padmanabhan,
  arXiv:hep-th/0607240.

\bibitem{Pad}T. Padmanabhan, Class. Quant. Grav. {\bf 19}, 5387
(2002)[arXiv: gr-qc/0204019]; Phys. Rept. {\bf 406}, 49
(2005)[arXiv: gr-qc/0311036]; T.~Padmanabhan,
  arXiv:gr-qc/0606061.

\bibitem{a5} G. W. Gibbons and S. W. Hawking, Phys. Rev. D\textbf{15}, 2738
(1977).
\bibitem{a6} R. G. Cai, Nucl. Phys. \textbf{B628}, 375(2002) [arXiv: hep-th/0112253];
R. G. Cai,
 Phys. Lett. \textbf{B525}, 331(2002)[arXiv:hep-th/0111093].

 \bibitem{a8} T. Jacobson, Phys. Rev. Lett. \textbf{75}, 1260(1995)[arXiv:
hep-th/0212327].

\bibitem{a11} R. G. Cai and S. P. Kim, JHEP
\textbf{0502}, 050 (2005)[arXiv:hep-th/0501055]
 \bibitem{a12} M.
Akbar and R. G. Cai, Phys. Lett. B\textbf{635}, 7 (2006).
[arXiv:hep-th/0602156]

\bibitem{a7} A. V. Frolov and L. Kofman, JCAP \textbf{0305}, 009
(2003).
\bibitem{a9} U. K. Danielsson, Phys. Rev.
D\textbf{71}, 023516(2005)[arXiv: hep-th/0411172].
 \bibitem{a10} R. Bousso, Phys. Rev. D\textbf{71}, 064024(2005)[arXiv:
hep-th/0412197].

 \bibitem{ain}G.~Calcagni,
  JHEP {\bf 0509}, 060 (2005)
  [arXiv:hep-th/0507125].

\bibitem{a23}C.W. Misner and D. H. Sharp, Phys. Rev. \textbf{136},
B571 (1964).

\bibitem{a15} S. A. Hayward, S. Mukohyana, and M. C. Ashworth, Phys. Lett. A\textbf{256},
347(1999); S.~A.~Hayward,
  Class.\ Quant.\ Grav.\  {\bf 15}, 3147 (1998)
  [arXiv:gr-qc/9710089].

\bibitem{a14} D. Bak and S. J. Rey, Class. Quant. Grav.
\textbf{17}, L83 (2000).


\bibitem{wang}B.~Wang, Y.~Gong and E.~Abdalla,
  arXiv:gr-qc/0511051.
\bibitem{a16} R. M. Wald, Phys. Rev. D\textbf{48}, 3427(1993).

\bibitem{ain2}D.~G.~Boulware and S.~Deser,
  Phys.\ Rev.\ Lett.\  {\bf 55}, 2656 (1985);
J.~T.~Wheeler,
  Nucl.\ Phys.\ B {\bf 268}, 737 (1986);
  Nucl.\ Phys.\ B {\bf 273}, 732 (1986);
R.~C.~Myers and J.~Z.~Simon,
  Phys.\ Rev.\ D {\bf 38}, 2434 (1988).

\bibitem{a17} R. G. Cai, Phys. Rev. D\textbf{65}, 084014 (2002);
 R. G. Cai and Q. Guo, Phys. Rev. D\textbf{69}, 104025 (2004).


\bibitem{a19} R. G. Cai and Y. S. Myang, Phys. Lett. B\textbf{559},
60(2003).
\bibitem{a20} D. Lovelock, J. Math. Phys. \textbf{12}, 498 (1971).

\bibitem{a22} R. G. Cai, Phys. Lett.
B\textbf{582}, 237(2004)[arXiv: hep-th/0311240].

\bibitem{others} M.~Banados, C.~Teitelboim and J.~Zanelli,
  Phys.\ Rev.\  D {\bf 49}, 975 (1994)
  [arXiv:gr-qc/9307033];
R.~G.~Cai and K.~S.~Soh,
  Phys.\ Rev.\  D {\bf 59}, 044013 (1999)
  [arXiv:gr-qc/9808067];
 J.~Crisostomo, R.~Troncoso and J.~Zanelli,
  Phys.\ Rev.\  D {\bf 62}, 084013 (2000)
  [arXiv:hep-th/0003271];
 R.~Aros, R.~Troncoso and J.~Zanelli,
  Phys.\ Rev.\  D {\bf 63}, 084015 (2001)
  [arXiv:hep-th/0011097].



\bibitem{Padd} A.~Mukhopadhyay and T.~Padmanabhan,
  arXiv:hep-th/0608120.

\bibitem{ACC}M. Akbar and R.G. Cai, gr-qc/0612089; R.G. Cai and L.M
Cao, gr-qc/0611071; R.G. Cai and L.M. Cao, hep-th/0612144.





\end{thebibliography}
\end{document}